\documentclass[journal=ancac3, manuscript=article]{achemso}

\usepackage{graphicx}
\usepackage{epstopdf}
\usepackage{longtable}
\usepackage{color}

\usepackage{dcolumn}
\usepackage{bm}

\author{Qian Jia}
\affiliation[Renmin University of China]{Department of Physics, Renmin University of China, Beijing 100872, China}

\author{Wei Ji}
\affiliation[Renmin University of China]{Department of Physics, Renmin University of China, Beijing 100872, China}
\alsoaffiliation[McGill University]{Centre for the Physics of Materials and Department of Physics, McGill University, 3600 rue University, Montreal, Canada H3A 2T8}
\email{wji@ruc.edu.cn}

\author{Sarah A. Burke}
\affiliation[The University of British Columbia]{Department of Physics and Astronomy, and the Department of Chemistry, The University of British Columbia, 6224 Agricultural Road, Vancouver, Canada V6T 1Z1}
\alsoaffiliation[McGill University]{Centre for the Physics of Materials and Department of Physics, McGill University, 3600 rue University, Montreal, Canada H3A 2T8}
\email{sburke@ubc.ca}

\author{Hong-Jun Gao}
\affiliation[Chinese Academy of Sciences]{Institute of Physics, Chinese Academy of Sciences, PO Box 603, Beijing, 100190 China}

\author{Peter Gr\"utter}
\affiliation[McGill University]{Centre for the Physics of Materials and Department of Physics, McGill University, 3600 rue University, Montreal, Canada H3A 2T8}

\author{Hong Guo}
\affiliation[McGill University]{Centre for the Physics of Materials and Department of Physics, McGill University, 3600 rue University, Montreal, Canada H3A 2T8}

\title{Adsorption of PTCDA and C$_{60}$ on KBr(001): electrostatic interaction versus electronic hybridization}


\begin{document}

\begin{abstract}

The adsorption of functional molecules on insulator surfaces is of great importance to molecular electronics. Interplay between these molecules and insulator substrates is still not very clear. In this work, we present a systematical investigation of geometric and electronic properties of perylene-3,4,9,10-tetracarboxylic-3,4,9,10-dianhydride (PTCDA) and C$_{60}$ on KBr(001) using density functional theory and non-contact atomic force microscopy. After energetic and structural details being discussed, electronic structures, e.g. local electronic density of states, (differential) charge density, and Bader charge analysis, were inspected. It was found that electrostatics is the primary interaction mechanism for PTCDA and C$_{60}$ adsorbed on KBr, which can be further promoted by electronic hybridizations of non-polar molecules, e.g. C$_{60}$, and the substrate. The electronic hybridization was suggested depending on the polarizability of the $\pi$-system, which can be most likely further suppressed by introducing high electron affinity atoms, \emph{e.g.} O, into the molecule. The adsorption site of a molecule on an ionic crystalline surface is, therefore, predominantly determined by electrostatics. Internal charge redistributions at the molecule-insulator interfaces was found for the both systems. Since the interaction mechanism is predominated by electrostatics, it was thus concluded that alkali-halides is a competitive candidate to be adopted to support low polarizability molecules, such as PTCDA, in future electronics. The influence of the found internal charge variations on electronic transport properties is still an open question, which may inspire new research fields of electronics.

\end{abstract}


\section{Introduction}
Using molecules as functional elements in electronic devices is a very interesting concept that has attracted a great deal of attention for several decades\cite{Ratner1974, Molecular-Devices, Organic-Materials, Mol-metal1, Mol-metal2, Mol-metal3, nature2000,Mohn2010}. While tremendous progress has been realized, it is recognized that molecular electronics is a very complicated problem as not only isolated molecules must be understood, but also their assembly, interaction, and contact to the outside world, must be carefully investigated\cite{Molecular-Devices, Organic-Materials, Mol-metal1, Mol-metal2, Mol-metal3, nature2000}. Since practically viable molecular circuitry is most likely to be self-assembled on solid substrates that are electrically insulated, a critical issue is how such a molecular layer is influenced by the underlying insulating surfaces. As demonstrated by many scanning tunneling microscopy (STM) studies\cite{Mol-metal1,Mol-metal2,Mol-metal3}, metal and semiconductor surfaces, most of which are conducting, substantially influence the electronic structure of molecules.

Recently, ultrathin insulator films were successfully fabricated on metal surfaces \cite{pent-nacl,c60-charging,Mura2010,Such2008}, which allows STM to image molecules on these ultrathin films. Non-contact atomic force microscopy (NC-AFM) was also demonstrated a feasible tool to investigate molecules on insulator surfaces \cite{Qi2011,HighQ}. Increasingly more experimental investigations have thus been focused on insulator surfaces or ultrathin films, \emph{e.g.} KBr(001)\cite{burke2005,burke2007,nanolett2004,ptcda-kbr,Pakarinen2009} or NaCl(001)\cite{pent-nacl,c60-charging,Dewetting,dewetting1,dewetting3,LocalContactPot}. However, only a few efforts have been made on the role of insulator surfaces to molecular overlayers, especially for electronic structures. Insulator surfaces were expected electronic inert to molecular overlayers, nevertheless, it has not yet been reported a clear physical picture of how molecule-substrate electronic hybridization is suppressed on insulators. In addition, the way that how molecules interact with insulator surfaces is also lack of full understanding.

\begin{figure}[pbt]
\includegraphics[width=8.6cm]{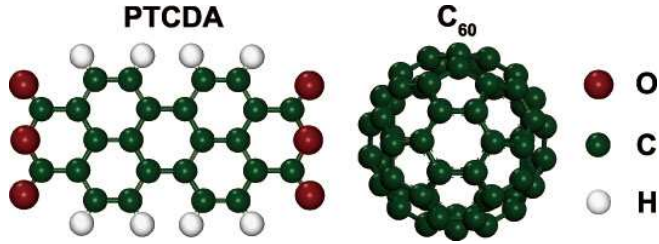}%
\caption{Ball-stick model showing the atomic structures of PTCDA (left) and C$_{60}$ (right)}
\label{fig:mols}
\end{figure}

It is therefore the purpose of this work endeavoring to improve the understanding of these issues. In this work, we have systematically investigated the geometric and electronic structures of C$_{60}$ and
perylene-3,4,9,10-tetracarboxylic-3,4,9,10-dianhydride (PTCDA) molecules (see Fig. \ref{fig:mols}) adsorbed on KBr(001) using density functional theory calculations and NC-AFM. The KBr (001) surface is one of the most
widely adopted insulator substrates for investigating molecular
overlayers\cite{burke2005,burke2007,nanolett2004,ptcda-kbr}. A polar molecule, {\it i.e.} PTCDA, and a nonpolar one, {\it i.e.} C$_{60}$, were considered, since they were widely investigated and are representative molecules in molecular electronics\cite{Organic-Materials, Mol-metal1, Mol-metal2, Mol-metal3}. Meanwhile, experimental literature of C$_{60}$ and PTCDA adsorbed on KBr(001) were available\cite{burke2005,burke2007,ptcda-kbr,Dewetting,dewetting1,dewetting3} for our results to compare with.

In particular, DFT calculations with the dispersion correction (DFT-D)\cite{DFT-D2001,dft-d,Grimme2011} were adopted to find the most likely adsorption configuration for each system and compared with our NC-AFM experiments. Adsorption energy and structural details were discussed. Given the adsorption configuration, electronic structures, e.g. local density of states (LDOS), were calculated to analysis the details of electronic hybridization, from which electrostatics was suggested the primary interaction between the molecules and the substrate. It was also found that the highest occupied molecular orbitals (HOMO) of C$_{60}$ electronically hybridizes with the Br {\it p}-state of the KBr substrate, forming a few new states; while there is no appreciable molecule-substrate electronic hybridization found in PTCDA/KBr, but a collective shift of molecular orbitals higher than HOMO, owing to a bending of PTCDA after adsorption. Further proofs, including differential charge density (DCD), real space distribution of wavefunctions, and structural distortions, were examined, all of which support the electrostatics. By investigating these two distinctly different molecules on a widely adopted substrate KBr, our results shed considerable light on improving the knowledge of the interplay of functional molecules and insulator surfaces. Furthermore, we inferred several general features for the influences of insulator surfaces to molecular overlayers by comparing C$_{60}$/KBr with PTCDA/KBr, which were expected to assist in choosing appropriate molecules and substrates for molecular nanoelectronics.

\section{Computational details}
Density functional theory calculations were carried out using the standard DFT-Projector-Augmented-Wave (DFT-PAW) method\cite{paw} with Perdew-Burke-Ernzerhof (PBE) functional\cite{pbe} and a its revised form of RPBE functional for exchange-correlation energy and planewave basis sets with energy cutoff up to 400 eV, as implemented in the Vienna ab initio Simulation Package (VASP) \cite{vasp,vasp2}. Semi-core \textit{p} electrons of K were treated as valance electrons. Five alkali-halide layers, separated by a vacuum slab equivalent to seven alkali-halide layer thickness (23\AA), were employed to model the surface. The two bottom layers were kept fixed during structural relaxations and all other atoms were fully relaxed until the net force on every relaxed atom is less than 0.01 eV/\AA. A p($2\times3$) and a p($3\times3$) supercell were  employed to model the PTCDA/KBr monolayer and a p($4\times4$) one to approximately simulate the (8$\times$3) monolayer\cite{burke2005} of C$_{60}$ on KBr. The surface Brillouin Zone of both categories of supercells were sampled by a $2\times2\times1$ $k$-point grid.

A dispersion correction in the form of DFT-D2 \cite{dft-d} was applied to PBE (PBE-D) and RPBE (RPBE-D) functionals, respectively. It was demonstrated that the RPBE-D method is the most suitable one for modeling molecule-metal interfaces among PBE, PBE-D, and RPBE-D\cite{lan-1t}. It is expected that RPBE-D is also the most suitable one for molecule-insulator interfaces. In the rest of the paper, we thus focus on the results of RPBE-D while that of PBE were also reported for comparison.

\section{Results and Discussion}

\subsection{Atomic Structures}
\subsubsection{PTCDA/KBr(001)}

\begin{figure}[tbp]
\includegraphics[width=8.6cm]{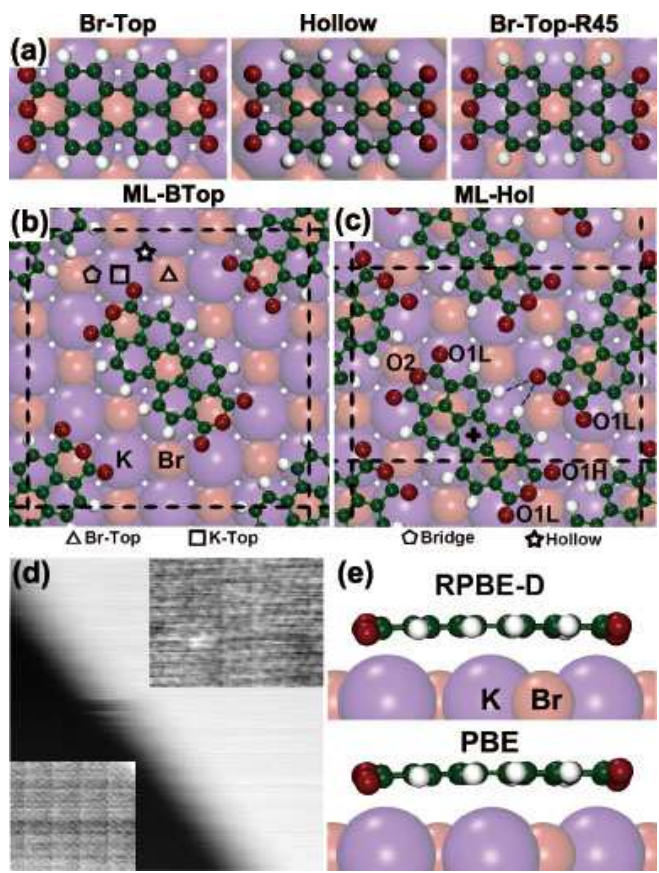}%
\caption{Top views of the RPBE-D method optimized structures in single-molecule configurations (a) and in monolayer configurations (b) and (c), in which the two supercells are marked by the black dashed lines. The K and Br atoms are represented by larger purple and smaller brown spheres, respectively. The ``triangle'', ``square'', ``pentagon'', and ``star'' represent the adsorption sites right on top of a Br (Br-Top) and a K (K-Top) atom, in between of a K and its adjacent Br and at the center of the square made by two Br and two K atoms. The center of a PTCDA is indicated by a ``+'' in (c) and O1L, O1H, and O2 refer to lower- and higher-carboxylic O and anhydride O, respectively. Two blue thin dashed lines in (c) shows likely hydrogen bonds between two PTCDAs. A corresponding NC-AFM image to configuration ML-Hol was shown in (d). Side views of PBE and RPBE-D fully relaxed monolayer structures (ML-Hol) are shown in (e).}
\label{fig:ptcda-br}
\end{figure}

Recent literature indicates that the carboxylic-O-cation interaction primarily contributes to the molecule-substrate interaction for PTCDA adsorbed on alkali-halide surfaces\cite{burke2008,ptcda-nacl,burke2009,Dewetting,dewetting1,dewetting3,LocalContactPot}. We thus considered adsorption sites where the carboxylic-oxygens (denote O1) are near K cations. In terms of a single PTCDA adosrbed on KBr, three adsorption sites were selected among four adsorption sites and two adsorption orientations. They are Br-Top, Hollow, and Br-Top-R45, as shown in Fig. \ref{fig:ptcda-br} (a). Total energy calculations show that the Br-Top site is the most stable, with an adsorption energy of 1.30 eV (0.60 eV for the PBE value), while that of the other two configurations are 0.80 (0.20) eV for Hollow and 0.86 (0.26) eV for Br-Top-R45.

The situation changes when PTCDA molecules form a monolayer. The smallest monolayer supercell that adsorption site Br-Top can form is a (3$\times$3) one, but a smaller (2$\times3$) one for the Hollow. A Smaller supercell means higher coverage and stronger intermolecular interaction, which may lead to configuration Br-Top, the most preferred one for single-molecule, unfavored in a monolayer. Two monolayer configurations were considered, {\it i.e.}, the Br-Top site (the most favored among all the single-molecule configurations) and the Hollow site(the most compact among all the monolayer configurations), denoted as ML-BTop and ML-Hol , respectively (Fig. \ref{fig:ptcda-br}(b) and (c)). Since ML-BTop is a (3$\times$3) supercell, its results, e.g. adsorption energy, were renormalized to a (2$\times$3) supercell in order to compare with that of ML-Hol. Results of adsorption energy prefer the ML-Hol configuration, ascribed to a stronger intermolecular interaction. The adsorption energy of ML-Hol,  -1.27 eV (per molecule, the same hereinafter), is 0.34 eV more stable than the ML-BTop value of -0.93 eV. Those values calculated using PBE are -0.58 eV, 0.14 eV, and -0.44 eV, respectively. Configuration ML-Hol is therefore the theoretical suggested monolayer configuration.

Figure \ref{fig:ptcda-br} (d) shows a NC-AFM image of PTCDA/KBr, which was acquired at the edge of a PTCDA island. The atomically resolved KBr surface and molecularly resolved PTCDA molecules are also visible in that image. It clearly suggests a (2$\times$3) superstructure where PTCDA molecules are along the [110] directions and the center of a PTCDA was assessed at a Hollow site. The shape of the supercell, the molecular orientation, and the adsorption site observed by the experiment are highly consistent with our theoretical results, which compellingly indicates configuration ML-Hol the most favorable structure of PTCDA/KBr(001).

In configuration ML-Hol, the four O1 atoms of a PTCDA are on top of four K cations and the central phenyl ring of the molecule is at the hollow site of two Br anions and two K cations. From the top view of the ML-Hol, all C atoms of a PTCDA can be assigned into two categories, {\it i.e.} two arm-chair edges that contain 16 C atoms being on top of K cations (Over-K C atoms) and two C-C bonds, comprised of four C atoms, in the middle of a PTCDA where C atoms are over Br anions (Over-Br C atoms). Figure \ref{fig:ptcda-br}(e) shows the side view of a fully relaxed PTCDA in configuration ML-Hol. The relaxed structure using either RPBE-D or PBE indicates a bent PTCDA, in which the vertical positions of the four O1 atoms are slightly lower than that of the almost planar perylene core, similar to PTCDA/Ag(111)\cite{ptcda,ptcda1,PTCDA-Ag1,PTCDA-Ag2}. As PTCDA is a symmetric molecule, the four O1 atoms shall be identical. When these molecules aggregate into a monolayer, a symmetry breaking was, however, found that the four O1 atoms diagonally split into two categories according to Table \ref{tab:ads-ptcda-ML}. One category is 0.1 \AA~vertically higher than the other. We denote the higher one as O1H and O1L for the lower one, as show in Fig. \ref{fig:ptcda-br}(c). Three angles are available to reflect the bending of PTCDA, namely angles $\alpha$1, $\alpha$2, and $\alpha$3. They are 174.8$^{\circ}$,  180.2$^{\circ}$, and  178.3$^{\circ}$, respectively, suggesting a clear bending of the molecule. The larger value of angle $\alpha$3, nearly  180$^{\circ}$, stems from the higher vertical position of O1H than that of O1L, which is most likely ascribed to the two hydrogen bonds (bond length of 2.82~\AA~and 2.92~\AA) formed between an O1H and two H atoms of a adjacent PTCDA, marked by the thinner blue dashed lines in Fig. \ref{fig:ptcda-br}(c).

Table \ref{tab:ads-ptcda-ML} shows the details of calculated molecule-substrate distances by RPBE-D and PBE. In terms of RPBE-D results, $d_{O1L-K}$, representing the averaged distance between a O1L and the K cation underneath, is 3.36 \AA, which is 0.20\AA~lower than that of O1H (3.56 \AA). Both distances are substantially shorter than the sum of vdW radii of O (1.52\AA) and K (2.75\AA), suggesting the molecule-substrate interaction shall be stronger than vdW interactions, which was further supported by the comparison between RPBE-D and PBE. The PBE distances, in which vdW interactions were not considered, only 0.2-0.3~\AA~longer than that of RPBE-D, still significantly shorter than the vdW radii sum. The qualitatively agreement between the two methods indicates that other interaction mechanisms, rather than the vdW interaction, are predominate for PTCDA adsorbed on KBr, as elucidated later.

\begin{table*}
\caption{\label{tab:ads-ptcda-ML}Adsorption energy (E$_{ads}$) and structural details of configuration ML-Hol calculated using PBE and RPBE-D. Angles $\alpha$1, $\alpha$2, and $\alpha$3 represent angles O1L-Ctr-O1L, O2-Ctr-O2, and O1H-Ctr-O1H, respectively, in which the ``Ctr" refers to the center of a PTCDA. Distances $d_{O1L-K}$, $d_{O1H-K}$, and $d_{Ctr-surf}$ indicate the distances between O1L and K underneath, O1H and K underneath, and from the center of a PTCDA to the substrate surface, respectively.}
\begin{tabular}{cccccccc}
\hline
\hline
 ML-Hol & $E_{ads}$ & $\alpha_{1}$ & $\alpha_{2}$
& $\alpha_{3}$ & $d_{O1L-K}$ & $d_{O1H-K}$ & $d_{Ctr-surf}$\\ \hline
 PBE    & -1.152~eV & 174.3$^{\circ}$ & 179.8$^{\circ}$ & 179.4$^{\circ}$& 3.54{\AA} & 3.87{\AA} & 3.85{\AA}

 \\
 RPBE-D & -2.532~eV & 174.8$^{\circ}$ & 180.2$^{\circ}$ & 178.3$^{\circ}$ & 3.36{\AA}& 3.56{\AA} &
 3.65{\AA}\\
 \hline
 \hline
\end{tabular}
\end{table*}

\subsubsection{C$_{60}$/KBr(001)}
\begin{figure}[bp]
\includegraphics[scale=0.6]{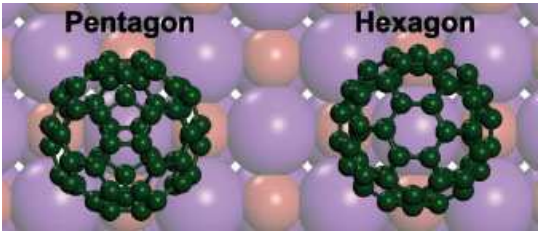}%
\caption{Top views of the PBE and (R)PBE-D suggested most stable configurations, {\it i.e.} C$_{60}$-DB (left) and C$_{60}$-Hex (right).} \label{fig:c60-kbr}
\end{figure}

Totally 18 configurations of C$_{60}$/KBr were considered based on four adsorption sites of the KBr surface, three contact positions of C$_{60}$ , {\it i.e.} pentagon, hexagon, double-bond, and two molecular orientations. Two configurations, revealed by both PBE and RPBE-D, show significant stability than the rest, by at least 0.1 eV. In different from that of PTCDA/KBr, PBE and RPBE-D suggest two distinct most stable configurations of C$_{60}$/KBr, respectively. Figure \ref{fig:c60-kbr} left and right show the most stable configurations revealed by PBE and RPBE-D, respectively. In the most PBE favored configuration, denoted C$_{60}$-DB, a double bond, equivalently a $\sim10^{\circ}$ rotated pentagon, faces to the K cation underneath; while in C$_{60}$-Hex, the most RPBE-D preferred configuration, a hexagon of the C$_{60}$ contacts to the K cation underneath.

Another method, {\it i.e.} PBE-D that the dispersion correction was applied to PBE functional, was adopted to clarify the issue of which method is more reliable. Table \ref{tab:ads-c60} shows the results of adsorption energy for C$_{60}$-DB and C$_{60}$-Hex. Although PBE shows 0.02 eV more stable of C$_{60}$-DB than C$_{60}$-Hex, either of the other two methods indicates a more stable C$_{60}$-Hex, 0.07 eV for PBE-D and 0.01 eV for RPBE-D. Since the dispersion corrected adsorption energy seems more reliable, it was believed that configuration C$_{60}$-Hex is more stable than C$_{60}$-DB, which was also confirmed by a NC-AFM experiment.

\begin{table}[tpb]
\caption{\label{tab:ads-c60}
The PBE, RPBE-D and PBE-D results of the adsorption energies of C$_{60}$-DB and C$_{60}$-Hex and their differences.}
\begin{tabular}{cccc}
\hline
\hline
\textrm{Adsorption Energy}&
\textrm{C$_{60-DB}$}&
\textrm{C$_{60-Hex}$}&
\textrm{C$_{60-Hex}$-C$_{60-DB}$}\\
\hline
  PBE~(eV) & -0.04  & -0.02 & 0.02 \\
  RPBE-D~(eV) &-0.39  & -0.40 & -0.01 \\
  PBE-D~(eV) & -0.56  & -0.63& -0.07\\
\hline
\hline
\end{tabular}
\end{table}

\begin{figure}[bp]
\includegraphics[ scale=1]{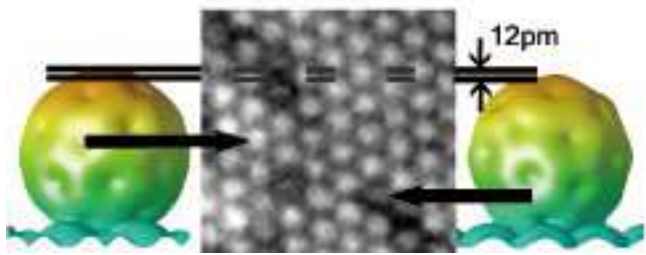}%
\caption{Isosurfaces (0.01~{\it e}/\AA$^{3}$) of the calculated total charge densities of C$_{60}$-Hex (left) and C$_{60}$-DB (right), in comparison with a NC-AFM image. The apparent height of the isosurface of C$_{60}$-Hex is at least 12 pm higher than that of C$_{60}$-DB, which is comparable to the experimental observed ``bright" and ``dim" spots (see middle). The colors of the isosurface are mapped to the height in the z direction (from surface towards vacuum).} \label{fig:afm}
\end{figure}

The contrast of NC-AFM images principally reflects of short and long range Coulomb interactions between sample and tip\cite{ncafm}, therefore, the corrugation of real-space total charge density of the sample can represent the height profile of NC-AFM measurements as a first order approximation. Figure \ref{fig:afm} shows the the calculated total charge densities of these two configurations. Since the molecule may freely rotate on the substrate, the apparent height of C$_{60}$-DB shall be determined by the lowest C-C bond of the top pentagon, which is 12~pm lower than the vertical height of C$_{60}$-Hex. Comparing with NC-AFM data, shown in Fig. \ref{fig:afm} middle, the height difference of 23~pm\cite{burke2005}, as well as the predominance of the ``bright" molecules corresponding to the more stable C$_{60}$-Hex configuration, with comparatively few ``dim" molecules corresponding to the lower and less energetic C$_{60}$-DB, are consistent with the theoretical findings here. In configuration C$_{60}$-Hex, the averaged C-K bond length of the lowest hexagon of C$_{60}$ is 3.53 \AA, and that of the double bond in configuration C$_{60}$-DB is 3.28 \AA. Both bond lengths, clearly shorter than the sum of the vdW radii of K and C, also implies a stronger interaction mechanism, rather than the vdW interaction, as elucidated below.

\subsection{Electronic structure}
Local partial density of states (LPDOS) of PTCDA/KBr (in configuration ML-Hol) and C$_{60}$/KBr (in configuration C$_{60}$-Hex) were calculated using the RPBE-D method which gives essentially the same results to the PBE functional, as shown in Figs. \ref{fig:es-ptcda} and \ref{fig:es-c60}, respectively. All energies were referenced to the vacuum level in both figures. Real space distribution of the HOMO and the lowest unoccupied molecular orbital (LUMO) of PTCDA and C$_{60}$ after adsorption were also discussed in this subsection.

\subsubsection{Non-hybridized PTCDA on KBr}
\begin{figure}[bp]
\includegraphics[scale=0.4]{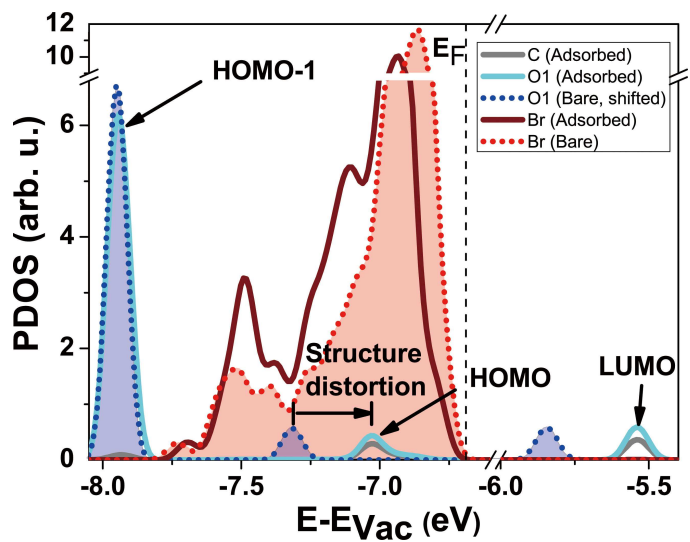}%
\caption{Local partial density of states for a carboxylic oxygen (O1) and an ``averaged'' C atom in PTCDA, and a Br anion under Over-Br C atoms in PTCDA/KBr. All energies are reference to the Vacuum Level (hereinafter). All states of the bare molecule and the bare substrate were plotted using dotted lines with shadows, while the states after adsorption are all in solid lines. Some molecular orbitals are indicated by the black arrows. The ``E$_{F}$'' here refers to the energy of the highest occupied state.}
\label{fig:es-ptcda}
\end{figure}

Figure \ref{fig:es-ptcda} shows the LPDOS of O1 (carboxylic O), C, and Br atoms of PTCDA/KBr before (dotted lines) and after (solid lines) the adsorption. It was found that the gap between HOMO-1 and HOMO of an adsorbed PTCDA is 0.3 eV larger than that of a bare PTCDA, but other gaps between adjacent upper MOs, e.g. HOMO-LUMO gap, do not change. Such a collective shift of MOs was ascribed to the adsorption induced structural distortion, \textit{i.e.} a bent ($\sim6.4^{\circ}$ for RPBE-D) molecular backbone, as supported by a calculation of a bent bare PTCDA, which implies that PTCDA does not appear significantly influenced by the KBr substrate. In particular, the molecular states of PTCDA do not hybridize to substrate states in any appreciable way. The HOMO of the adsorbed PTCDA situates at roughly 0.3~eV lower than the upper-edge of the Br {\it p} band ($\sim$~-6.7~eV), not aligning to the band edge. The highest energy level of this band edge does not appreciably move, within several meV, before and after the adsorption. All these results suggest that the charge transfer between PTCDA and KBr is very small, in other words, no covalent bond was formed between the molecule and the substrate, which was confirmed by the differential charge density and Bader charge analysis discussed later.

\subsubsection{Hybridization between C$_{60}$ and KBr}
\begin{figure}[tbp]
\includegraphics[scale=0.4]{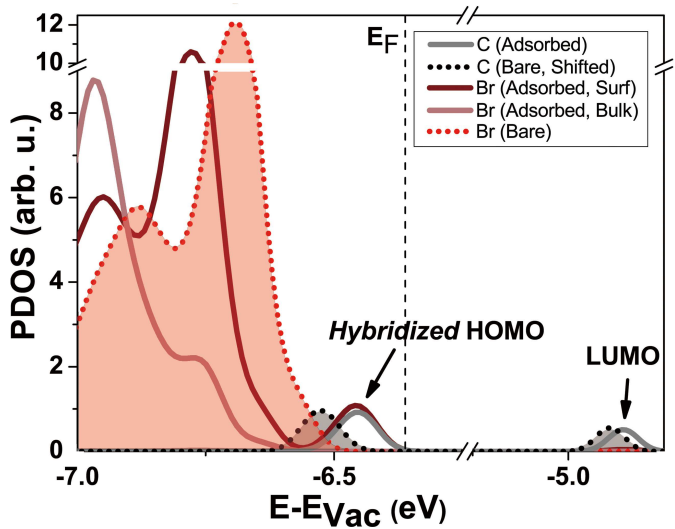}%
\caption{Local partial density of states of a C atom in the bottom hexagon layer of  C$_{60}$ and
a Br  anion among the four Br anions under the molecule in C$_{60}$/KBr. The styles of figure drawing are the same to those of Fig. \ref{fig:es-ptcda}.}
\label{fig:es-c60}
\end{figure}

Partial local DOSs of a C atom in the bottom layer of C$_{60}$ and one of the four Br anions under C$_{60}$ before and after the adsorption of C$_{60}$ on KBr were plotted in Fig. \ref{fig:es-c60}. It unambiguously shows a C-Br hybridized state, in contrast to PTCDA/KBr where no molecule-substrate electronic hybridization was found. The hybridized HOMO state, sitting at -6.45~eV, is comprised of molecular HOMO and a surface Br state, whereas bulk states of Br (light brown solid line) do not contribute to it. The adsorption of C$_{60}$ lowers the work function of bare KBr(001) by 0.1 eV, resulting in an upward shift of the upper edge of the occupied states. The HOMO-LUMO gaps between adsorbed (gray solid line) and bare (black dotted line) C$_{60}$ are slightly different, which is due to the upward shift of C$_{60}$'s HOMO. These results, combined with the slightly downward shifted edge of the surface Br $p$ band, suggest a charge transfer from KBr to C$_{60}$, consistent with the electron-acceptor nature of C$_{60}$.

\subsubsection{Real space distribution of electronic states}
\begin{figure}[bpt]
\includegraphics[width=8.6cm]{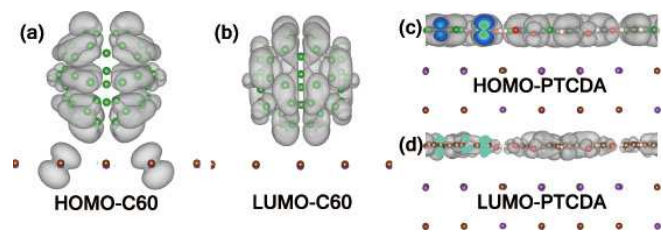}%
\caption{Real space distribution of the wavefunctions of HOMO (a) and LUMO (b) of C$_{60}$/KBr(001); and HOMO (c) and LUMO (d) of PTCDA/KBr(001). } \label{fig:states}
\end{figure}

Electronic states near band gaps are of particular interest for molecular electronics. We thus plotted the real space distribution (RSD) of a few states originating from HOMO and LUMO of C$_{60}$ and PTCDA, to illustrate what was found above, \emph{i.e.} electronic hybridization between C$_{60}$ and KBr and the electronically inert PTCDA on KBr. Figure \ref{fig:states}(a) and (b) show the RSDs of the hybridized HOMO residing at -6.45~eV and a C$_{60}$'s LUMO locating at -4.86 eV, respectively. The -6.45 eV state is the hybridized state, consistent with the LPDOSs. Interestingly, the LUMO of C$_{60}$ does not hybridize with the substrate, hence it keeps its original shape and energy. On the other hand, as shown in Fig. \ref{fig:states}(c) and (d), neither PTCDA's HOMO nor its LUMO electronically interact with KBr, again, consistent with the non-hybridized LPDOSs.

\subsubsection{Strength of hybridization}
An interesting question has thus arisen that why does C$_{60}$ hybridize with KBr but PTCDA does not? A key difference between PTCDA and C$_{60}$ is that in PTCDA, the O atoms have higher electron affinity than C atoms, which significantly increases the polarizability of the rest of the $\pi$-electron system (perylene core) of PTCDA by drawing electrons towards the ends of the molecule. The enhanced polarizability of $\pi$-electrons suppresses the ability of electronic hybridization, leading to the less reactive C atoms rather than O atoms for frontier MOs\cite{ptcda}. Nevertheless, O atoms in PTCDA still have possibility to hybridize with substrate Br, forming covalent-like bonds. However, since both of Br and O are of high electron affinity, they are thus negatively charged leading to a repulsive electrostatic interaction rather than an attractive one, which makes the electronic hybridization between them unlikely. However, the negatively charged Br anions and/or O atoms and positively charged K cation give rise to a predominantly electrostatic interaction instead. These findings suggest that the polarizability of molecules largely determines the strength of electronic hybridization between molecules and alkali-halides.

\subsection{Interaction mechanism}
Interaction mechanism between the adsorbate(s) and the substrate is of great importance to an adsorption. In this subsection, two electrostatic interaction mechanisms, distinguished by whether it is electronic hybridization enhanced, were revealed for PTCDA/KBr and C$_{60}$/KBr, respectively, according to differential charge density, Bader change analysis, and structural distortions upon adsorption. All the calculations in this subsection were performed with the RPBE-D method if not specified.

\subsubsection{Differential charge density}
\begin{figure}[tpb]
\includegraphics[width=8.6 cm]{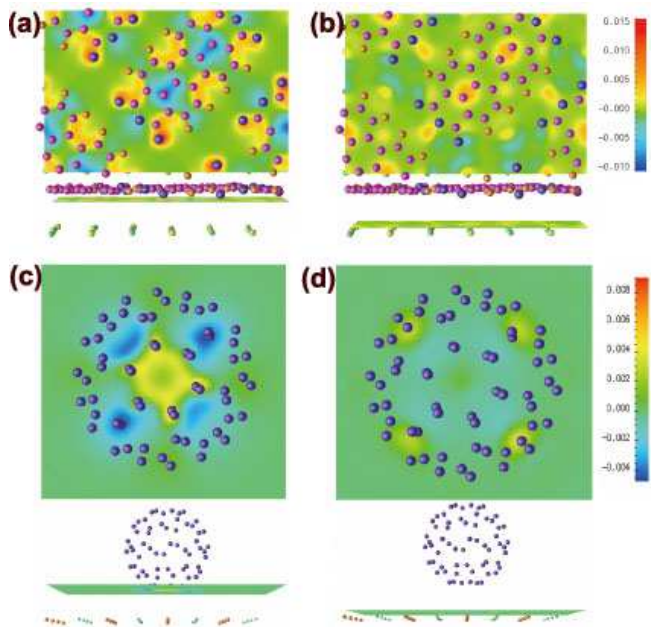}%
\caption{Top views (upper) and side views (lower) of differential charge densities (DCD) of PTCDA/KBr(001) in slabs near the molecules (a) and close to the substrate (b); and those of C$_{60}$/KBr(001) in slabs just below the molecule (c) and just above the substrate (d). The colors are mapped by the DCD values, in unit of $e$/\AA$^{3}$, as illustrated by the color bars.}%
\label{fig:dcd}%
\end{figure}

Differential charge density (DCD) is defined by $\rho_{DCD}=\rho_{Total}-\rho _{Molecule}- \rho_{Substrate}$. Figure \ref{fig:dcd}(a) and (b) show the DCDs of PTCDA at the two slabs illustrated in the associated lower panels, respectively. A charge accumulation (warn colors) around the O1 and Over-K C atoms and a charge reduction (cold colors) of the O2 and Over-Br C atoms were explicitly shown in Fig. \ref{fig:dcd} (a) where the charge density slab is slightly below PTCDA molecules. An opposite tendency was found in the DCD of a slab close to the substrate (Fig. \ref{fig:dcd}(b)), \textit{i.e.}, a slight charge reduction of the surface K cations below O1 atoms and a significant charge accumulation around the Br anions under O2 and Over-Br C atoms. In a slab near C$_{60}$ (Fig. \ref{fig:dcd} (c)), a large amount of charge accumulation was found around the lowest hexagon of C$_{60}$ which is over a K cation of KBr, while slight charge reductions were observable at the higher C atoms. In terms of the slab right above the substrate (Fig. \ref{fig:dcd}(d)), a tiny charge accumulation is appreciable over the said K cation (at the center of the panel) and relatively larger charge accumulations appear on the Br anions adjacent to the said K cation.

All the DCDs share the same feature that charge accumulations and reductions are vertically superposed for both PTCDA and C$_{60}$ on KBr, which unambiguously suggests an electrostatic interaction mechanism for the molecule-substrate interaction. In terms of PTCDA/KBr, such mechanism was concluded the primarily interaction for the interface, since there was no appreciable electronic hybridization between PTCDA and KBr. Although electronic hybridization was found in C$_{60}$/KBr, it induces a charge redistribution, largely enhancing the strength of electrostatic interactions. The electrostatic interaction is, therefore, still a very important part for the interface interaction. Bader charge analysis and the resulted surface structural distortions support these statements.

\subsubsection{Bader charge analysis}

\begin{figure}[pbt]
\includegraphics[width=8.2 cm]{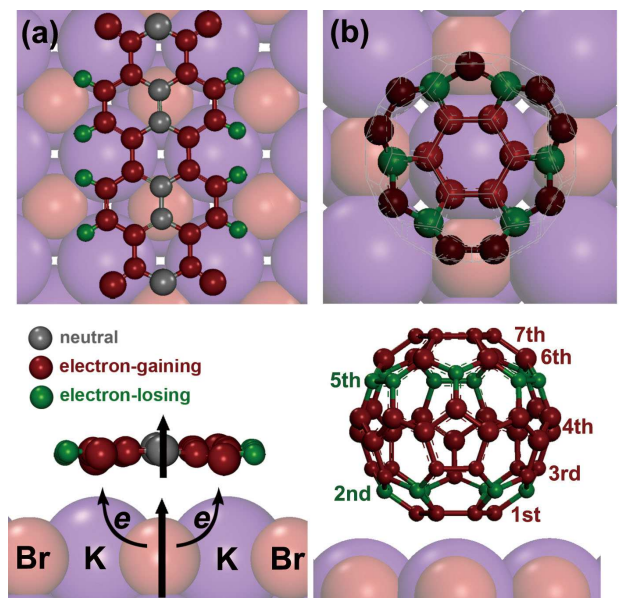}%
\caption{Sketches of charge variations of PTCDA (a) and C$_{60}$ (b) on KBr before and after the adsorption. Green, gray, and red balls represent the electron-losing, neutral, and electron-gaining atoms after the adsorption. In term of (a), O atoms are shown by larger balls and C atoms are in small balls. Two electric dipole moments were marked on the lower panel by the two black arrows and the Br-K-PTCDA charge transfer was illustrated by the two thinner black arrows associated with an ``{\it e}'' for each of the both. Only the bottom three layers of C$_{60}$ were shown in balls in the upper panel of (b), while in the lower panel, C$_{60}$ was divided into seven lateral layers, from the first to the seventh.}%
\label{fig:bader}%
\end{figure}

Bader charge analysis was employed to quantitatively investigate the DCD-suggested charge redistribution and likely molecule-substrate charge transfer. Figure \ref{fig:bader}(a) illustrated the charge redistribution of PTCDA/KBr, in which electrons relocate from the H atoms (green balls) to the Over-K C atoms, and the middle C and O2 atoms are essentially neutral. Table \ref{tab:bader} listed the charge variations of different categories of atoms that these ``green balls" totally loses a charge of 0.22 {\it e} per molecule and a charge of 0.30 {\it e} is gained by those ``red balls". An extra charge of 0.08 {\it e} was donated from the substrate, primarily surface Br anions, to PTCDA.

\begin{table}[pbt]
\caption{\label{tab:bader}
Charge variations of different categories of atoms for PTCDA on KBr calculated by the Bader charge analysis. ``Layer 1'' refers to the interfacial layer of KBr (hereinafter).}
\begin{tabular}{cccccccc}
\hline
\hline
   & C$_{ctr}$ & C$_{edge}$& O1& O2 & H & K  & Br     \\  \hline
 PTCDA ({\it e})   &+0.01&+0.26& +0.04& -0.01 & -0.22 & - & -\\
 Layer 1 ({\it e})  &- &-& -& - & - & +0.02 & -0.10  \\
 Layer 2 ({\it e})  &- &-& -& - & - & +0.00 & -0.00   \\
 Layer 3 ({\it e})  &- &-& -& - & - & +0.00 & -0.00   \\
 Layer 4 ({\it e})  &- &-& -& - & - & +0.02 & -0.01   \\
 Layer 5 ({\it e})  &- &-& -& - & - & +0.02 & -0.03   \\
  Total & \multicolumn{5}{c}{+0.08}  & \multicolumn{2}{c}{-0.08}  \\
\hline
\hline
\end{tabular}
\end{table}

These results of Bader charge analysis looks a bit inconsistent with the DCDs, however, it provides more information if considering both results together. It can be inferred that the electrons surrounding the Over-Br C and O2 atoms were pushed up from the interstitial space between them and the surface Br anions to the above, due to the molecule-Br Coulomb repulsion, which forms electric dipoles normal to the surface, as shown in Fig. \ref{fig:bader}(a). To response these dipoles, imaging dipoles were also induced at Br anions underneath. In other words, the charge redistribution among these atoms, {\it i.e.} Over-Br C, O2, and Br atoms, is perpendicular to the surface plane. Such molecule-Br Coulomb repulsion also gives rise to a in-plane charge transfer from Br anions to their adjacent surface K cations. Less positive K cations is against electrostatic interaction between K cations and Br anions. A straightforward solution is a charge transfer from the K cations to the PTCDA above, which strengthens both the in-plane K-Br and the normal-to-plane molecule-K interactions. This hypothesis was confirmed by the fact that a charge of 0.08~{\it e} was transferred from the substrate Br anions to a PTCDA. A in-plane charge redistribution within PTCDA was also found that electrons transfer from H atoms to Over-K C or O1 atoms. They are all consistent with the electrostatic interaction mechanism that electric dipoles at Over-Br C atoms and Br anions, the K-mediated charge transfer from Br anions to PTCDA, negative charges on Over-K C and O1 atoms that over K cations, and additional positive charges on H atoms that close to Br anions underneath and negatively charged Over-K-C and O1 atoms.

The amount of transferred electrons from KBr to C$_{60}$ is half to that of PTCDA, as shown in Table \ref{tab:bader-c60}. Charge redistribution was found throughout the molecule and the substrate, except the bottom layer of the substrate slab. The four Br anions under C$_{60}$ loses a charge of 0.04~{\it e}, while the all surface Br anions gain, overall, a charge of 0.01~{\it e} and a charge of 0.04~{\it e} was lost for all surface K cations. The most significant difference from the results of PTCDA/KBr is that a charge of up to $\sim$ 0.1~{\it e} was gained by Br anions from K cations in each bulk layer of KBr, which gives rise to more stable bulk layers. In terms of C$_{60}$, the first and third layers that contact to the substrate (red balls in Fig. \ref{fig:bader}(b)) gain charges of 0.05 {\it e} and 0.04~{\it e}, respectively, while the second layer (green balls), in between those two layers gaining electrons, loses a charge of 0.04~{\it e}. The alternating appearance of electron gains and losses, also found for the fourth to sixth layers, lowers the total energy of C$_{60}$. In addition, the transferred electrons to the ``red" C atoms of the first layer attract with the K cation underneath, further strengthening the molecule-substrate interaction. The attraction aside, it also reduces the molecule-substrate (Br-C) Coulomb repulsion that the electron losses of the second layer C atoms (0.04~{\it e}) and the adjacent Br anions (0.04~{\it e}), which considerably stabilizes the interface. These results explicitly suggest a charge transfer from the four Br anions, not from the surface K cations, to the C$_{60}$, which is consistent with the found C-Br electronic hybridization. All these facts manifest that electrostatic interaction, promoted by electronic hybridizations, is primarily responsible for the C$_{60}$-KBr interaction, accordant with the DCD results.

\begin{table*}[pbtp]
\caption{\label{tab:bader-c60}
Charge variations of different categories of atoms for C$_{60}$/KBr revealed by the Bader charge analysis. The ``1'' to ``7'' in column C$_{60}$ refer to the layers illustrated in Fig. \ref{fig:bader}(b) and the ``1'' and ``5'' in column KBr represent the first and bottom layers of the KBr slab, respectively.}
\tabcolsep 0.05in
\begin{tabular}{ccccccccccccccc}
\hline
\hline
 & \multicolumn{7}{c}{C$_{60}$ ({\it e})} & \multicolumn{5}{c}{KBr ({\it e})} & Total  \\ \cline{2-8}
\cline{9-14}
 & 1 & 2 & 3 & 4 & 5 & 6 & 7 & 1 & 2 & 3 & 4 & 5 & \\ \hline
 C    &+0.05& -0.04 & +0.04 & +0.10 & -0.24 & +0.10 & +0.03  & - & - & - & - &- & +0.04\\
 K  & - & - & -& - & - & -& - & -0.04 & -0.09 & -0.08 &  -0.03 & 0.00 & -0.24\\
 Br & - & - & -& - & - & -& - & +0.01 & +0.07 & +0.10 & +0.02 & 0.00 & +0.20 \\
\hline
\hline
\end{tabular}
\end{table*}

\subsubsection{Structural distortion of the substrate} 
\begin{figure}[tpb]
\includegraphics[width=8.6 cm]{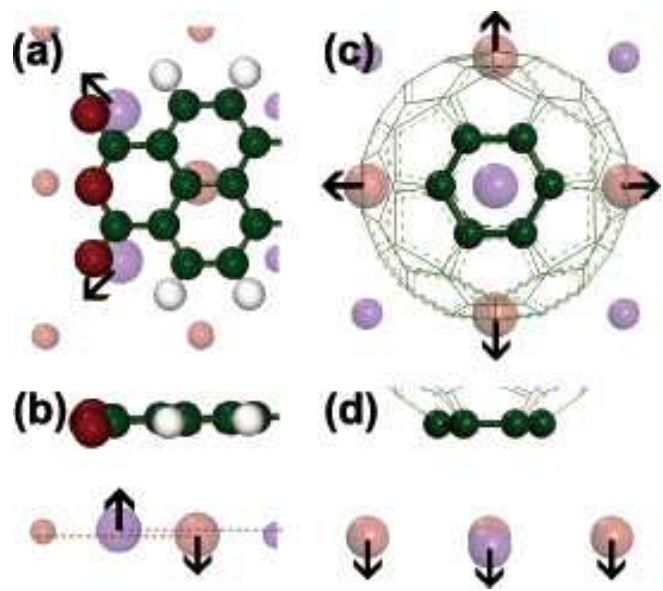}%
\caption{Structural distortions of K cations and Br anions at the interface during the adsorption of PTCDA (a) and (b) and C$_{60}$ (c) and (d) in top views (a) and (c) and side views (b) and (d). All larger balls, together with black arrows, were considered in discussion. The direction of these black arrows indicates the direction of movement for the considered atoms.}%
\label{fig:distortion}%
\end{figure}

Figure \ref{fig:distortion} illustrated the molecule-adsorption-induced structural distortions of KBr. In terms of PTCDA/KBr, the two negatively charged O1 atoms of a half PTCDA, shown in Fig. \ref{fig:distortion}(a), effectively attract the two K cations underneath, leading to the K cations being pulled out from their initial positions by 0.06~\AA~ along the directions indicated by the black arrows in Fig. \ref{fig:distortion}(a). Vertical distortions were also found for K cations and Br anions that the two K cations under O1 are vertically pulled up by 0.10~\AA (0.09~\AA~ for PBE), as shown in Fig. \ref{fig:distortion}(b). The Br anion under the C-C bond appears tend to be away from the surface, since it is less negative, {\it i.e.} less attraction between it and the K cation underneath after the adsorption. However, considering the electron densities of both the Br anion and the C atoms are rather high, a repulsion between them does exist, which results in the Br anion being overall pushed down by 0.05~\AA (0.02~\AA~ for PBE). Both are significant vertical shifts compared to any anticipated rumpling of the KBr(001)\cite{rumpling}.

Similar to the PTCDA case, The repulsion between Br anions and C atoms pushes these four Br anion under the C$_{60}$ outwards, as indicated by the black arrows in Fig. \ref{fig:distortion}(c). While a substantial charge redistribution was found for C$_{60}$/KBr, significant vertical distortions are expected for K cations and Br anions. Figure \ref{fig:distortion}(d) shows that these four Br anions are vertically lowered by 0.08~\AA~ (0.06~\AA~ for PBE), which is, again, ascribed to the repulsion between Br anions and C atoms. The more positive K cation beneath C$_{60}$ moves downwards by 0.06~\AA~ (0.09~\AA~ for PBE), owing to a weaker attraction between it and the surrounding Br anions whose lose a charge of 0.04 {\it e}. All these distortions provide further proofs that electrostatics should be responsible for the primary interaction of this interface.

\section{Conclusion}
In summary, we have determined the most likely configuration for PTCDA/KBr(001), namely the O-over-K$^{+}$ configuration, and a flat-lying hexagon and a tilted pentagon configuration for that of C$_{60}$/KBr(001). On the basis of these configurations, electrostatics was revealed the primary interaction mechanism for PTCDA and C$_{60}$ adsorbed on KBr, which can be further strengthened by electronic hybridizations of non-polar molecules, e.g. C$_{60}$. The electronic hybridization between halogens and C $\pi$-systems depends on the polarizability of  the $\pi$-system, which can be most likely further suppressed by introducing high electron affinity atoms, \emph{e.g.} O, into the $\pi$-system. Internal charge redistributions, especially vertical ones, were also found at the molecule-substrate interface for both PTCDA and C$_{60}$. These charge redistributions can be regarded as a response to the periodic potential generated by Br anions and K cations at the KBr surface. Additional conclusions are in order, (a) appreciable electronic hybridization can be found for C$_{60}$/KBr; (b) the LUMO and upper MOs for both systems, as well as the HOMO for PTCDA/KBr, are almost uninfluenced by the insulating substrate; (c) the adsorption site of isolated molecules on ionic crystalline surfaces is predominantly determined by electrostatics.

Due to the dominant electrostatic interaction mechanism, we conclude that alkali-halides is a competitive candidate to be adopted to support low polarizability molecules such as PTCDA in future electronics. In terms of high polarizability molecules, like C$_{60}$, side groups or other introduced high electron affinity atoms were expected to help preventing or even tuning the hybridization. It would be thus interesting to examine the role of alkali-halides in other purely carbon based $\pi$-systems, like graphene nano ribbons, and the role of substrate-induced internal charge redistributions in electron transport properties in future studies.

\begin{acknowledgement}
We gratefully acknowledge financial support by the National Natural Science Foundation of China (NSFC, Grant No. 11004244 and 11274380), the Beijing Natural Science Foundation (BNSF, Grant No. 2112019), and the Basic Research Funds in Renmin University of China from the Central Government (Grant No. 12XNLJ03) (W.J.); ......for S.A.B.; the NSERC of Canada, FQRNT of Quebec and CIFAR (H.G. and P.G.). W.J. was supported by the Program for New Century Excellent Talents in University. We are grateful to Mr. Zhi-Xin Hu for his technical assistance for DFT calculations and the Physics Lab for High-Performance Computing of RUC and Shanghai Supercomputing Center for substantial computer time.
\end{acknowledgement}

\bibliography{acsnano-mol-kbr}

\providecommand*\mcitethebibliography{\thebibliography}
\csname @ifundefined\endcsname{endmcitethebibliography}
  {\let\endmcitethebibliography\endthebibliography}{}
\begin{mcitethebibliography}{41}
\providecommand*\natexlab[1]{#1}
\providecommand*\mciteSetBstSublistMode[1]{}
\providecommand*\mciteSetBstMaxWidthForm[2]{}
\providecommand*\mciteBstWouldAddEndPuncttrue
  {\def\EndOfBibitem{\unskip.}}
\providecommand*\mciteBstWouldAddEndPunctfalse
  {\let\EndOfBibitem\relax}
\providecommand*\mciteSetBstMidEndSepPunct[3]{}
\providecommand*\mciteSetBstSublistLabelBeginEnd[3]{}
\providecommand*\EndOfBibitem{}
\mciteSetBstSublistMode{f}
\mciteSetBstMaxWidthForm{subitem}{(\alph{mcitesubitemcount})}
\mciteSetBstSublistLabelBeginEnd
  {\mcitemaxwidthsubitemform\space}
  {\relax}
  {\relax}

\bibitem[Aviram and Ratner(1974)Aviram, and Ratner]{Ratner1974}
Aviram,~A.; Ratner,~M.~A. Molecular rectifiers. \emph{Chemical Physics Letters}
  \textbf{1974}, \emph{29}, 277 -- 283\relax
\mciteBstWouldAddEndPuncttrue
\mciteSetBstMidEndSepPunct{\mcitedefaultmidpunct}
{\mcitedefaultendpunct}{\mcitedefaultseppunct}\relax
\EndOfBibitem
\bibitem[Reed et~al.(2003)Reed, Lee, and Eds]{Molecular-Devices}
Reed,~M.; Lee,~T.; Eds, Nanoelectronics, M. 2003\relax
\mciteBstWouldAddEndPuncttrue
\mciteSetBstMidEndSepPunct{\mcitedefaultmidpunct}
{\mcitedefaultendpunct}{\mcitedefaultseppunct}\relax
\EndOfBibitem
\bibitem[Fraxedas(2006)]{Organic-Materials}
Fraxedas,~J. Molecular organic materials. 2006\relax
\mciteBstWouldAddEndPuncttrue
\mciteSetBstMidEndSepPunct{\mcitedefaultmidpunct}
{\mcitedefaultendpunct}{\mcitedefaultseppunct}\relax
\EndOfBibitem
\bibitem[Barlow and Raval(2003)Barlow, and Raval]{Mol-metal1}
Barlow,~S.; Raval,~R. Complex organic molecules at metal surfaces: bonding,
  organisation and chirality. \emph{Surface Science Reports} \textbf{2003},
  \emph{50}, 201 -- 341\relax
\mciteBstWouldAddEndPuncttrue
\mciteSetBstMidEndSepPunct{\mcitedefaultmidpunct}
{\mcitedefaultendpunct}{\mcitedefaultseppunct}\relax
\EndOfBibitem
\bibitem[Rosei et~al.(2003)Rosei, Schunack, Naitoh, Jiang, Gourdon, Laegsgaard,
  Stensgaard, Joachim, and Besenbacher]{Mol-metal2}
Rosei,~F.; Schunack,~M.; Naitoh,~Y.; Jiang,~P.; Gourdon,~A.; Laegsgaard,~E.;
  Stensgaard,~I.; Joachim,~C.; Besenbacher,~F. Properties of large organic
  molecules on metal surfaces. \emph{Progress in Surface Science}
  \textbf{2003}, \emph{71}, 95 -- 146\relax
\mciteBstWouldAddEndPuncttrue
\mciteSetBstMidEndSepPunct{\mcitedefaultmidpunct}
{\mcitedefaultendpunct}{\mcitedefaultseppunct}\relax
\EndOfBibitem
\bibitem[Papageorgiou et~al.(2004)Papageorgiou, Salomon, Angot, Layet,
  Giovanelli, and Lay]{Mol-metal3}
Papageorgiou,~N.; Salomon,~E.; Angot,~T.; Layet,~J.-M.; Giovanelli,~L.;
  Lay,~G.~L. Physics of ultra-thin phthalocyanine films on semiconductors.
  \emph{Progress in Surface Science} \textbf{2004}, \emph{77}, 139 -- 170\relax
\mciteBstWouldAddEndPuncttrue
\mciteSetBstMidEndSepPunct{\mcitedefaultmidpunct}
{\mcitedefaultendpunct}{\mcitedefaultseppunct}\relax
\EndOfBibitem
\bibitem[Joachim et~al.(2000)Joachim, Gimzewski, and Aviram]{nature2000}
Joachim,~C.; Gimzewski,~J.~K.; Aviram,~A. Electronics using hybrid-molecular
  and mono-molecular devices. \emph{Nature} \textbf{2000}, \emph{408}, 541 --
  548\relax
\mciteBstWouldAddEndPuncttrue
\mciteSetBstMidEndSepPunct{\mcitedefaultmidpunct}
{\mcitedefaultendpunct}{\mcitedefaultseppunct}\relax
\EndOfBibitem
\bibitem[Mohn et~al.({2010})Mohn, Repp, Gross, Meyer, Dyer, and
  Persson]{Mohn2010}
Mohn,~F.; Repp,~J.; Gross,~L.; Meyer,~G.; Dyer,~M.~S.; Persson,~M. {Reversible
  Bond Formation in a Gold-Atom-Organic-Molecule Complex as a Molecular
  Switch}. \emph{{PHYSICAL REVIEW LETTERS}} \textbf{{2010}}, \emph{{105}}\relax
\mciteBstWouldAddEndPuncttrue
\mciteSetBstMidEndSepPunct{\mcitedefaultmidpunct}
{\mcitedefaultendpunct}{\mcitedefaultseppunct}\relax
\EndOfBibitem
\bibitem[Repp et~al.(2005)Repp, Meyer, Stojkovi\ifmmode~\acute{c}\else
  \'{c}\fi{}, Gourdon, and Joachim]{pent-nacl}
Repp,~J.; Meyer,~G.; Stojkovi\ifmmode~\acute{c}\else \'{c}\fi{},~S.~M.;
  Gourdon,~A.; Joachim,~C. Molecules on Insulating Films: Scanning-Tunneling
  Microscopy Imaging of Individual Molecular Orbitals. \emph{Phys. Rev. Lett.}
  \textbf{2005}, \emph{94}, 026803\relax
\mciteBstWouldAddEndPuncttrue
\mciteSetBstMidEndSepPunct{\mcitedefaultmidpunct}
{\mcitedefaultendpunct}{\mcitedefaultseppunct}\relax
\EndOfBibitem
\bibitem[Nazin et~al.(2005)Nazin, Qiu, and Ho]{c60-charging}
Nazin,~G.~V.; Qiu,~X.~H.; Ho,~W. Charging and Interaction of Individual
  Impurities in a Monolayer Organic Crystal. \emph{Phys. Rev. Lett.}
  \textbf{2005}, \emph{95}, 166103\relax
\mciteBstWouldAddEndPuncttrue
\mciteSetBstMidEndSepPunct{\mcitedefaultmidpunct}
{\mcitedefaultendpunct}{\mcitedefaultseppunct}\relax
\EndOfBibitem
\bibitem[Mura et~al.({2010})Mura, Gulans, Thonhauser, and
  Kantorovich]{Mura2010}
Mura,~M.; Gulans,~A.; Thonhauser,~T.; Kantorovich,~L. {Role of van der Waals
  interaction in forming molecule-metal junctions: flat organic molecules on
  the Au(111) surface}. \emph{{PHYSICAL CHEMISTRY CHEMICAL PHYSICS}}
  \textbf{{2010}}, \emph{{12}}, {4759--4767}\relax
\mciteBstWouldAddEndPuncttrue
\mciteSetBstMidEndSepPunct{\mcitedefaultmidpunct}
{\mcitedefaultendpunct}{\mcitedefaultseppunct}\relax
\EndOfBibitem
\bibitem[Such et~al.({2008})Such, Goryl, Godlewski, Kolodziej, and
  Szymonski]{Such2008}
Such,~B.; Goryl,~G.; Godlewski,~S.; Kolodziej,~J.~J.; Szymonski,~M. {PTCDA
  molecules on a KBr/InSb system: a low temperature STM study}.
  \emph{{NANOTECHNOLOGY}} \textbf{{2008}}, \emph{{19}}\relax
\mciteBstWouldAddEndPuncttrue
\mciteSetBstMidEndSepPunct{\mcitedefaultmidpunct}
{\mcitedefaultendpunct}{\mcitedefaultseppunct}\relax
\EndOfBibitem
\bibitem[Qi({2011})]{Qi2011}
Qi,~Y. {Investigation of organic films by atomic force microscopy: Structural,
  nanotribological and electrical properties}. \emph{{SURFACE SCIENCE REPORTS}}
  \textbf{{2011}}, \emph{{66}}, {379--393}\relax
\mciteBstWouldAddEndPuncttrue
\mciteSetBstMidEndSepPunct{\mcitedefaultmidpunct}
{\mcitedefaultendpunct}{\mcitedefaultseppunct}\relax
\EndOfBibitem
\bibitem[LeDue et~al.({2009})LeDue, Lopez-Ayon, Burke, Miyahara, and
  Grutter]{HighQ}
LeDue,~J.~M.; Lopez-Ayon,~M.; Burke,~S.~A.; Miyahara,~Y.; Grutter,~P. {High Q
  optical fiber tips for NC-AFM in liquid}. \emph{{NANOTECHNOLOGY}}
  \textbf{{2009}}, \emph{{20}}\relax
\mciteBstWouldAddEndPuncttrue
\mciteSetBstMidEndSepPunct{\mcitedefaultmidpunct}
{\mcitedefaultendpunct}{\mcitedefaultseppunct}\relax
\EndOfBibitem
\bibitem[Burke et~al.(2005)Burke, Mativetsky, Hoffmann, and
  Gr\"utter]{burke2005}
Burke,~S.~A.; Mativetsky,~J.~M.; Hoffmann,~R.; Gr\"utter,~P. Nucleation and
  Submonolayer Growth of ${\mathrm{C}}_{60}$ on KBr. \emph{Phys. Rev. Lett.}
  \textbf{2005}, \emph{94}, 096102\relax
\mciteBstWouldAddEndPuncttrue
\mciteSetBstMidEndSepPunct{\mcitedefaultmidpunct}
{\mcitedefaultendpunct}{\mcitedefaultseppunct}\relax
\EndOfBibitem
\bibitem[Burke et~al.(2007)Burke, Mativetsky, Fostner, and
  Gr\"utter]{burke2007}
Burke,~S.~A.; Mativetsky,~J.~M.; Fostner,~S.; Gr\"utter,~P. ${\mathrm{C}}_{60}$
  on alkali halides: Epitaxy and morphology studied by noncontact AFM.
  \emph{Phys. Rev. B} \textbf{2007}, \emph{76}, 035419\relax
\mciteBstWouldAddEndPuncttrue
\mciteSetBstMidEndSepPunct{\mcitedefaultmidpunct}
{\mcitedefaultendpunct}{\mcitedefaultseppunct}\relax
\EndOfBibitem
\bibitem[Nony et~al.(2004)Nony, Gnecco, Baratoff, Alkauskas, Bennewitz,
  Pfeiffer, Maier, Wetzel, Meyer, and Gerber]{nanolett2004}
Nony,~L.; Gnecco,~E.; Baratoff,~A.; Alkauskas,~A.; Bennewitz,~R.; Pfeiffer,~O.;
  Maier,~S.; Wetzel,~A.; Meyer,~E.; Gerber,~C. Observation of Individual
  Molecules Trapped on a Nanostructured Insulator. \emph{Nano Letters}
  \textbf{2004}, \emph{4}, 2185--2189\relax
\mciteBstWouldAddEndPuncttrue
\mciteSetBstMidEndSepPunct{\mcitedefaultmidpunct}
{\mcitedefaultendpunct}{\mcitedefaultseppunct}\relax
\EndOfBibitem
\bibitem[Mativetsky et~al.(2007)Mativetsky, Burke, Fostner, and
  Grutter]{ptcda-kbr}
Mativetsky,~J.; Burke,~S.; Fostner,~S.; Grutter,~P. Templated growth of 3, 4,
  9, 10-perylenetetracarboxylic dianhydride molecules on a nanostructured
  insulator. \emph{Nanotechnology} \textbf{2007}, \emph{18}, 105303\relax
\mciteBstWouldAddEndPuncttrue
\mciteSetBstMidEndSepPunct{\mcitedefaultmidpunct}
{\mcitedefaultendpunct}{\mcitedefaultseppunct}\relax
\EndOfBibitem
\bibitem[Pakarinen et~al.({2009})Pakarinen, Mativetsky, Gulans, Puska, Foster,
  and Grutter]{Pakarinen2009}
Pakarinen,~O.~H.; Mativetsky,~J.~M.; Gulans,~A.; Puska,~M.~J.; Foster,~A.~S.;
  Grutter,~P. {Role of van der Waals forces in the adsorption and diffusion of
  organic molecules on an insulating surface}. \emph{{PHYSICAL REVIEW B}}
  \textbf{{2009}}, \emph{{80}}\relax
\mciteBstWouldAddEndPuncttrue
\mciteSetBstMidEndSepPunct{\mcitedefaultmidpunct}
{\mcitedefaultendpunct}{\mcitedefaultseppunct}\relax
\EndOfBibitem
\bibitem[Topple et~al.({2011})Topple, Burke, Ji, Fostner, Tekiel, and
  Gruetter]{Dewetting}
Topple,~J.~M.; Burke,~S.~A.; Ji,~W.; Fostner,~S.; Tekiel,~A.; Gruetter,~P.
  {Tailoring the Morphology and Dewetting of an Organic Thin Film}.
  \emph{{JOURNAL OF PHYSICAL CHEMISTRY C}} \textbf{{2011}}, \emph{{115}},
  {217--224}\relax
\mciteBstWouldAddEndPuncttrue
\mciteSetBstMidEndSepPunct{\mcitedefaultmidpunct}
{\mcitedefaultendpunct}{\mcitedefaultseppunct}\relax
\EndOfBibitem
\bibitem[Burke et~al.({2009})Burke, Topple, and Gruetter]{dewetting1}
Burke,~S.~A.; Topple,~J.~M.; Gruetter,~P. {Molecular dewetting on insulators}.
  \emph{{JOURNAL OF PHYSICS-CONDENSED MATTER}} \textbf{{2009}},
  \emph{{21}}\relax
\mciteBstWouldAddEndPuncttrue
\mciteSetBstMidEndSepPunct{\mcitedefaultmidpunct}
{\mcitedefaultendpunct}{\mcitedefaultseppunct}\relax
\EndOfBibitem
\bibitem[Burke et~al.({2008})Burke, Ji, Mativetsky, Topple, Fostner, Gao, Guo,
  and Grutter]{dewetting3}
Burke,~S.~A.; Ji,~W.; Mativetsky,~J.~M.; Topple,~J.~M.; Fostner,~S.;
  Gao,~H.~J.; Guo,~H.; Grutter,~P. {Strain induced dewetting of a molecular
  system: Bimodal growth of PTCDA on NaCl}. \emph{{PHYSICAL REVIEW LETTERS}}
  \textbf{{2008}}, \emph{{100}}\relax
\mciteBstWouldAddEndPuncttrue
\mciteSetBstMidEndSepPunct{\mcitedefaultmidpunct}
{\mcitedefaultendpunct}{\mcitedefaultseppunct}\relax
\EndOfBibitem
\bibitem[Burke et~al.({2009})Burke, LeDue, Miyahara, Topple, Fostner, and
  Grutter]{LocalContactPot}
Burke,~S.~A.; LeDue,~J.~M.; Miyahara,~Y.; Topple,~J.~M.; Fostner,~S.;
  Grutter,~P. {Determination of the local contact potential difference of PTCDA
  on NaCl: a comparison of techniques}. \emph{{NANOTECHNOLOGY}}
  \textbf{{2009}}, \emph{{20}}\relax
\mciteBstWouldAddEndPuncttrue
\mciteSetBstMidEndSepPunct{\mcitedefaultmidpunct}
{\mcitedefaultendpunct}{\mcitedefaultseppunct}\relax
\EndOfBibitem
\bibitem[X.~Wu and Scoles(2001)X.~Wu, and Scoles]{DFT-D2001}
X.~Wu,~S. N. V.~L.,~M. C.~Vargas; Scoles,~G. Towards extending the
  applicability of density functional theory to weakly bound systems. \emph{J.
  Chem. Phys.} \textbf{2001}, \emph{115}, 8748\relax
\mciteBstWouldAddEndPuncttrue
\mciteSetBstMidEndSepPunct{\mcitedefaultmidpunct}
{\mcitedefaultendpunct}{\mcitedefaultseppunct}\relax
\EndOfBibitem
\bibitem[Grimme(2006)]{dft-d}
Grimme,~S. Semiempirical GGA-type density functional constructed with a
  long-range dispersion correction. \emph{Journal of Computational Chemistry}
  \textbf{2006}, \emph{27}, 1787--1799\relax
\mciteBstWouldAddEndPuncttrue
\mciteSetBstMidEndSepPunct{\mcitedefaultmidpunct}
{\mcitedefaultendpunct}{\mcitedefaultseppunct}\relax
\EndOfBibitem
\bibitem[Grimme(2011)]{Grimme2011}
Grimme,~S. Density functional theory with London dispersion corrections.
  \emph{Wiley Interdisciplinary Reviews: Computational Molecular Science}
  \textbf{2011}, \emph{1}, 211--228\relax
\mciteBstWouldAddEndPuncttrue
\mciteSetBstMidEndSepPunct{\mcitedefaultmidpunct}
{\mcitedefaultendpunct}{\mcitedefaultseppunct}\relax
\EndOfBibitem
\bibitem[Bl\"ochl(1994)]{paw}
Bl\"ochl,~P.~E. Projector augmented-wave method. \emph{Phys. Rev. B}
  \textbf{1994}, \emph{50}, 17953--17979\relax
\mciteBstWouldAddEndPuncttrue
\mciteSetBstMidEndSepPunct{\mcitedefaultmidpunct}
{\mcitedefaultendpunct}{\mcitedefaultseppunct}\relax
\EndOfBibitem
\bibitem[Perdew et~al.(1996)Perdew, Burke, and Ernzerhof]{pbe}
Perdew,~J.~P.; Burke,~K.; Ernzerhof,~M. Generalized Gradient Approximation Made
  Simple. \emph{Phys. Rev. Lett.} \textbf{1996}, \emph{77}, 3865--3868\relax
\mciteBstWouldAddEndPuncttrue
\mciteSetBstMidEndSepPunct{\mcitedefaultmidpunct}
{\mcitedefaultendpunct}{\mcitedefaultseppunct}\relax
\EndOfBibitem
\bibitem[Kresse and Furthm¨¹ller(1996)Kresse, and Furthm¨¹ller]{vasp}
Kresse,~G.; Furthm¨¹ller,~J. Efficiency of ab-initio total energy calculations
  for metals and semiconductors using a plane-wave basis set.
  \emph{Computational Materials Science} \textbf{1996}, \emph{6}, 15 --
  50\relax
\mciteBstWouldAddEndPuncttrue
\mciteSetBstMidEndSepPunct{\mcitedefaultmidpunct}
{\mcitedefaultendpunct}{\mcitedefaultseppunct}\relax
\EndOfBibitem
\bibitem[Kresse and Joubert(1999)Kresse, and Joubert]{vasp2}
Kresse,~G.; Joubert,~D. From ultrasoft pseudopotentials to the projector
  augmented-wave method. \emph{Phys. Rev. B} \textbf{1999}, \emph{59},
  1758--1775\relax
\mciteBstWouldAddEndPuncttrue
\mciteSetBstMidEndSepPunct{\mcitedefaultmidpunct}
{\mcitedefaultendpunct}{\mcitedefaultseppunct}\relax
\EndOfBibitem
\bibitem[Lan and Ji(2012)Lan, and Ji]{lan-1t}
Lan,~H.; Ji,~W. Role of the dispersion correction in the modeling of
  molecule-metal interfaces: Thiophene adsorbed on Cu (100), Cu (110) and Cu
  (111). \emph{Arxiv preprint arXiv:1205.6065} \textbf{2012}, \relax
\mciteBstWouldAddEndPunctfalse
\mciteSetBstMidEndSepPunct{\mcitedefaultmidpunct}
{}{\mcitedefaultseppunct}\relax
\EndOfBibitem
\bibitem[Burke et~al.(2008)Burke, Ji, Mativetsky, Topple, Fostner, Gao, Guo,
  and Gr\"utter]{burke2008}
Burke,~S.~A.; Ji,~W.; Mativetsky,~J.~M.; Topple,~J.~M.; Fostner,~S.;
  Gao,~H.-J.; Guo,~H.; Gr\"utter,~P. Strain Induced Dewetting of a Molecular
  System: Bimodal Growth of PTCDA on NaCl. \emph{Phys. Rev. Lett.}
  \textbf{2008}, \emph{100}, 186104\relax
\mciteBstWouldAddEndPuncttrue
\mciteSetBstMidEndSepPunct{\mcitedefaultmidpunct}
{\mcitedefaultendpunct}{\mcitedefaultseppunct}\relax
\EndOfBibitem
\bibitem[Karacuban et~al.(2011)Karacuban, Koch, Fendrich, Wagner, and
  M{\"o}ller]{ptcda-nacl}
Karacuban,~H.; Koch,~S.; Fendrich,~M.; Wagner,~T.; M{\"o}ller,~R. PTCDA on Cu
  (111) partially covered with NaCl. \emph{Nanotechnology} \textbf{2011},
  \emph{22}, 295305\relax
\mciteBstWouldAddEndPuncttrue
\mciteSetBstMidEndSepPunct{\mcitedefaultmidpunct}
{\mcitedefaultendpunct}{\mcitedefaultseppunct}\relax
\EndOfBibitem
\bibitem[Burke et~al.(2009)Burke, Topple, and Gr\"utter]{burke2009}
Burke,~S.~A.; Topple,~J.~M.; Gr\"utter,~P. Molecular dewetting on insulators.
  \emph{Journal of Physics: Condensed Matter} \textbf{2009}, \emph{21},
  423101\relax
\mciteBstWouldAddEndPuncttrue
\mciteSetBstMidEndSepPunct{\mcitedefaultmidpunct}
{\mcitedefaultendpunct}{\mcitedefaultseppunct}\relax
\EndOfBibitem
\bibitem[Ji et~al.(2006)Ji, Lu, and Gao]{ptcda}
Ji,~W.; Lu,~Z.-Y.; Gao,~H. Electron Core-Hole Interaction and Its Induced Ionic
  Structural Relaxation in Molecular Systems under X-Ray Irradiation.
  \emph{Phys. Rev. Lett.} \textbf{2006}, \emph{97}, 246101\relax
\mciteBstWouldAddEndPuncttrue
\mciteSetBstMidEndSepPunct{\mcitedefaultmidpunct}
{\mcitedefaultendpunct}{\mcitedefaultseppunct}\relax
\EndOfBibitem
\bibitem[Ji et~al.(2006)Ji, Lu, and Gao]{ptcda1}
Ji,~W.; Lu,~Z.-Y.; Gao,~H. Electron Core-Hole Interaction and Its Induced Ionic
  Structural Relaxation in Molecular Systems under X-Ray Irradiation.
  \emph{Phys. Rev. Lett.} \textbf{2006}, \emph{97}, 246101\relax
\mciteBstWouldAddEndPuncttrue
\mciteSetBstMidEndSepPunct{\mcitedefaultmidpunct}
{\mcitedefaultendpunct}{\mcitedefaultseppunct}\relax
\EndOfBibitem
\bibitem[Hauschild et~al.({2010})Hauschild, Temirov, Soubatch, Bauer, Schoell,
  Cowie, Lee, Tautz, and Sokolowski]{PTCDA-Ag1}
Hauschild,~A.; Temirov,~R.; Soubatch,~S.; Bauer,~O.; Schoell,~A.; Cowie,~B.
  C.~C.; Lee,~T.~L.; Tautz,~F.~S.; Sokolowski,~M. {Normal-incidence x-ray
  standing-wave determination of the adsorption geometry of PTCDA on Ag(111):
  Comparison of the ordered room-temperature and disordered low-temperature
  phases}. \emph{{PHYSICAL REVIEW B}} \textbf{{2010}}, \emph{{81}}\relax
\mciteBstWouldAddEndPuncttrue
\mciteSetBstMidEndSepPunct{\mcitedefaultmidpunct}
{\mcitedefaultendpunct}{\mcitedefaultseppunct}\relax
\EndOfBibitem
\bibitem[Brumme et~al.({2011})Brumme, Neucheva, Toher, Gutierrez, Weiss,
  Temirov, Greuling, Kaczmarski, Rohlfing, Tautz, and Cuniberti]{PTCDA-Ag2}
Brumme,~T.; Neucheva,~O.~A.; Toher,~C.; Gutierrez,~R.; Weiss,~C.; Temirov,~R.;
  Greuling,~A.; Kaczmarski,~M.; Rohlfing,~M.; Tautz,~F.~S. et~al.  {Dynamical
  bistability of single-molecule junctions: A combined experimental and
  theoretical study of PTCDA on Ag(111)}. \emph{{PHYSICAL REVIEW B}}
  \textbf{{2011}}, \emph{{84}}\relax
\mciteBstWouldAddEndPuncttrue
\mciteSetBstMidEndSepPunct{\mcitedefaultmidpunct}
{\mcitedefaultendpunct}{\mcitedefaultseppunct}\relax
\EndOfBibitem
\bibitem[Morita et~al.(2002)Morita, Wiesendanger, and Meyer]{ncafm}
Morita,~S.; Wiesendanger,~R.; Meyer,~E. \emph{Noncontact atomic force
  microscopy}; Springer Verlag, 2002; Vol.~1\relax
\mciteBstWouldAddEndPuncttrue
\mciteSetBstMidEndSepPunct{\mcitedefaultmidpunct}
{\mcitedefaultendpunct}{\mcitedefaultseppunct}\relax
\EndOfBibitem
\bibitem[Vogt and Weiss(2002)Vogt, and Weiss]{rumpling}
Vogt,~J.; Weiss,~H. The structure of KBr(100) and LiF(100) single crystal
  surfaces: a tensor low energy electron diffraction analysis. \emph{Surface
  Science} \textbf{2002}, \emph{501}, 203 -- 213\relax
\mciteBstWouldAddEndPuncttrue
\mciteSetBstMidEndSepPunct{\mcitedefaultmidpunct}
{\mcitedefaultendpunct}{\mcitedefaultseppunct}\relax
\EndOfBibitem
\end{mcitethebibliography}

\end{document}